\documentclass[preprint, aps,prb]{revtex4}

\usepackage{graphicx}

\def\gs{\gtrsim}
\def\ls{\lesssim}
\def\be{\begin{equation}}
\def\en{\end{equation}}                  
\def\p{\partial} 
\newcommand{\bi}[1]{\mbox{\boldmath$#1$}}
\newcommand{\av}[1]{\langle{#1}\rangle}

\def\bea{\begin{equation}\begin{array}{rcl}}
\def\ena{\end{array}\end{equation}}

\def\asigma{\stackrel{\leftrightarrow}{\sigma}}

\def\q{{\footnotesize{\it q}}\kern -5pt {\footnotesize{\it q}}}
\def\k{{\footnotesize{\it k}}\kern -5pt {\footnotesize{\it k}}}
\def\seq{\sim \kern -12pt \lower 5pt \hbox{$\displaystyle =$}}
\def\nnabla{\nabla\kern-3.3mm\nabla}

\def\ge{> \kern -12pt \lower 5pt \hbox{$\displaystyle =$}}
\def\le{< \kern -12pt \lower 5pt \hbox{$\displaystyle =$}}
\def\gs{> \kern -12pt \lower 5pt \hbox{$\displaystyle{\sim}$}}
\def\ls{< \kern -12pt \lower 5pt \hbox{$\displaystyle{\sim}$}}

\def\be{\begin{equation}}
\def\bea{\begin{eqnarray}}
\def\en{\end{equation}}
\def\ena{\end{eqnarray}}
\def\p{\partial }
 
\renewcommand{\theequation}{\arabic{section}.\arabic{equation}}

\begin{document}
\draft
\bibliographystyle{prsty}

\title{Dislocation Formation 
in Two-Phase Alloys}

\author{ Akihiko Minami and Akira Onuki}
\address{
Department of Physics, 
Kyoto University,  Kyoto 606-8502,  Japan}

\date{\today}

\begin{abstract}
A phase field model 
is presented to study  dislocation formation 
(coherency loss)  in two-phase binary alloys. 
In our model the   elastic energy density 
 is a  periodic function of the shear and tetragonal 
strains, which allows multiple 
formation of dislocations.  The composition 
is coupled to the elastic field twofold 
via  lattice misfit and via  composition-dependence 
of the elastic moduli. 
 By numerically integrating 
the dynamic equations in two dimensions, 
we find that dislocations 
appear in pairs in the interface region and grow
 into slips.  One end of each slip  glides   preferentially 
into the softer region, 
while the other end 
remains trapped  at the interface. 
Under uniaxial stretching at deep quenching,  
slips appear in the softer region 
and do not penetrate into the harder domains, 
giving rise to a gradual   increase of the stress 
 with increasing applied strain in  plastic 
flow.  
\end{abstract}

\pacs{62.20 Fe, 64.70Kb, 81.40 cd}
                             
\date{ }
\pagestyle{empty}

\maketitle

\section{Introduction}

In crystalline solids many kinds of phase transformations 
are strongly influenced by the elastic field 
\cite{Cahn71,Khabook,Fratzl,Onukibook}. 
Since the first work by Cahn \cite{Cahn1,Cahn2} 
most theoretical studies 
have been focused on the  coherent case in which 
the lattice planes are continuous through the 
interfaces.   In the incoherent case, on the other hand,  
 dislocations appear around the interfaces 
and the continuity is lost 
partially or even 
completely. 
Such incoherent  microstructures emerge 
  in various alloys  when  
 the lattice constants or 
the  crystalline structures 
of the two phases are not  close 
\cite{Strudel,Nembach,pollock,loss}. 
Moreover, they are produced  
in plastic flow because  dislocations 
generated tend to be trapped  at the interfaces. 
In particular, coherency loss has been 
extensivesly studied in the presence of 
$\gamma'$ precipitates 
of the  L1$_2$  structure \cite{pollock,loss}.

Theory for  the incoherent case is  much more 
difficult than for the coherent case, 
obviously because the effects cannot be 
adequately described within the usual linear elasticity 
theory \cite{Landau}.   The aim of this paper is hence 
to present a simple mathematical model 
reasonably describing  the incoherent effects in binary alloys. 
Use will be made 
of a recent nonlinear elasticity theory 
of plastic flow by one of the present authors 
\cite{OnukiPRE}.

A number of authors have studied 
composition changes  around 
dislocations fixed in space and time,  
which lead to  a compositional Cottrell 
atmosphere \cite{Cottrell} or preferential 
 nucleation around 
a dislocation \cite{CahnN,Tol}. 
As recent numerical work in two dimensions, 
phase separation has been studied by 
L{\'e}onard and Desai \cite{Desai} 
and by Hu and Chen \cite{Chen} 
using a continuum Ginzburg-Landau or phase field model
in the presence of fixed dislocations. 
In these papers, 
 dislocations preexist 
as singular objects before composition changes.   
We also mention atomistic simulations
of dislocation motion influenced by  diffusing 
solutes \cite{Sro} 
or by precipitated domains 
\cite{Chen3}.

Mechanical properties of two-phase solids are  
very different from those of one-phase 
solids \cite{Strudel,Cottrell,Haasen}. 
In the presence of precipitated  domains,  
dislocations can be pinned at the interface regions 
and networks of high-density dislocations 
can be formed preferentially in softer regions 
after deformations \cite{pollock}. 
These effects are very complex but important in technology.  
Our simulations will give some insights 
on the behavior of  dislocations in two-phase states.

This paper is organized as follows. In Section 2 
we will present the free energy functional 
for the composition and the elastic field, in which 
the  elastic energy density  is a periodic 
function of the tetragonal and shear strains 
and the composition is coupled to the elastic field.
In Section 3 we will construct 
 dynamical equations. 
In Section 4 numerical results will be 
given on the  
dislocation formation 
around domains and on the stress-strain relations 
under uniaxial stretching.

\section{Free Energy Functional}
\setcounter{equation}{0}

We consider  a  binary alloy 
consisting two components, $A$ and $B$, 
neglecting vacancies and interstitials. 
The compositions,  $c_A$ and $c_B$, 
of the two components satisfy  $c_A+c_B=1$. 
In real metallic alloys undergoing a phase transition,
there can be a change in 
the atomic configuration within unit cells 
as well as in the overall composition, resulting in 
ordered domains with the so-called 
L1$_0$  or  L1$_2$  structure 
\cite{Khabook,Onukibook,Desai1}. 
However, in this paper,  the composition difference 
is the   sole order parameter, 
\be 
\psi= c_A-c_B, 
\en
for simplicity.   The other variables representing 
the order-disorder phase transition are neglected. 
Then $\psi$ is  in the range $-1 \le \psi \le 1$ and 
\be 
c_A= \frac{1}{2}(1+\psi), \quad c_B= \frac{1}{2}(1-\psi).
\en

In our free energy  $F= \int d{\bi r} f$ 
 the order parameter 
$\psi$ and the elastic displacement 
vector ${\bi u}=(u_x,u_y)$ 
 are coupled. The free energy density $f$ is of the form, 
\begin{eqnarray}
f
=  f_{\rm BW}(\psi) + \frac{C}{2} | \nabla \psi |^2
 +\alpha e_1\psi +   f_{\rm{el}}   .  
\end{eqnarray}
The first term  is the Bragg-Williams 
free energy density expressed as \cite{Onukibook}  
\begin{eqnarray}
  \frac{v_0}{k_{\rm B}T}f_{\rm BW}  
&=& \frac{1 + \psi}{2}  \ln  ( 1 +
  \psi)  + \frac{1 - \psi}{2} 
\ln (1 - \psi)
\nonumber\\ &-& {T_{0}}\psi^2/2T ,  
\end{eqnarray}
where $v_0$ is the volume of a unit cell 
representing  the atomic volume, 
$ T_{0} $ is the mean-field critical temperature 
in the absence of the coupling to the elastic field. 
If $|\psi|\ll 1$, we obtain the Landau expansion 
$
{v_0} f_{\rm BW}/k_{\rm B}T 
=(1-{T_{0}}/T)\psi^2/2+  
\psi^4/24+\cdots.
$ 
However, we will not use 
this expansion form   because 
we are interested in 
the deeply quenched case. 
The second term in (2.3) 
is  the gradient term where 
 $C$ is  a positive constant. 
The parameter $\alpha$ 
represents the  strength of the coupling 
between the composition and the dilation strain $e_1=
 \nabla\cdot{\bi u}$. 
This coupling arises in the presence 
of a difference in the atomic sizes 
of  the two  species  
and is consistent with 
 the empirical fact that  the lattice constant 
changes linearly as a function of the 
average composition in many one-phase alloys 
(Vegard law). 
It gives rise to a 
difference in the 
lattice constants of the two phases 
 in phase separation (lattice misfit). It  
also explains  a composition  inhomogeneity 
(Cottrell atmosphere in one-phase states 
or  precipitate in two-phase states) 
around a dislocation.

In two dimensions $f_{\rm el}$ depends on 
the following  strains, 
\begin{eqnarray}
e_1 &=& \nabla _x u_x + \nabla _y u_y,\nonumber\\
e_2 &=& \nabla _x u_x - \nabla _y u_y,\nonumber\\
e_3 &=& \nabla _y u_x + \nabla _x u_y, 
\end{eqnarray}
where $\nabla_x= \p /\p x$ and $\nabla_y= \p /\p y$.  
The elastic displacement  ${\bi u}$ 
is measured in a reference  one-phase 
state at the critical composition. 
We call $e_2$ the tetragonal 
strain and $e_3$ the shear strain. 
In this paper we use a nonlinear elastic energy density  
of the form,  
\be 
 f_{\rm{el}} = \frac{1}{2}K  e_1^2 + 
{\Phi}(\psi, e_2, e_3 ). 
\en
The first term represents 
the elastic energy due to dilation with $K$ being the 
bulk modulus. The second term arises from 
anisotropic  shear deformations defined 
for arbitrary values of 
$e_2$ and $e_3$. Assuming a square lattice structure 
\cite{OnukiPRE},  we set 
\be
{\Phi}=\frac{\mu_2}{4\pi^2}[ 1 - \cos (2 \pi e_2)] 
+\frac{\mu_3}{4\pi^2} [ 1 - \cos (2 \pi e_3)]   .  
\en  
The principal crystal axes are 
 along  or make 
angles of $\pm \pi/4$ with respect to the $x$ or  $y$ axis. 
In Fig.1 we plot $\Phi$ as a function of 
$e_2$ and $e_3$ for the case 
$\mu_2=\mu_3=\mu_0$  in units of $\mu_0$. 
If the system is homogeneous, 
elastic stability is attained for $\p^2\Phi/\p e_2^2>0$ 
and   $\p^2\Phi/\p e_3^2>0$  or in the regions 
$|e_2-n|<1/4$ and 
 $|e_3-m|<1/4$ with $n$ and $m$ 
being integer values \cite{OnukiPRE}.

For small strains $|e_2| \ll 1$ and 
$|e_3| \ll 1$, it follows  the usual 
 standard  form \cite{Landau}, 
\be  
{\Phi}\cong  
\frac{1}{2}\mu_2 e_2^2 +\frac{1}{2}\mu_3 e_3^2, 
\en  
in the linear elasticity theory. Therefore,   
\be 
\mu_2= \frac{1}{2} (C_{11}-C_{12}), \quad \mu_3=C_{44}, 
\en 
in terms of the usual elastic moduli  
$C_{11}$, $C_{12}$, and $C_{44}$ \cite{Landau}.  
In the original theory \cite{Cahn1} the isotropic 
linear elasticity with constant 
$\mu_2=\mu_3$ was assumed. 
Subsequent theories treated 
the case of the cubic linear elasticity 
with  constant 
$\mu_2$ and $\mu_3$   
\cite{Khabook,Fratzl,Onukibook,Cahn2,Nishimori}. 
In the present paper, while $K$ is a constant,  
$\mu_2$ and $\mu_3$ depend on 
the composition as 
\be 
\mu_2 = 
\mu_{20} + \mu_{21} \psi, \quad   
\mu_3 = 
\mu_{30} + \mu_{31} \psi.  
\en 
If  $\mu_{21}>0$ and $\mu_{31}>0$, 
 the regions with larger (smaller) $\psi$ 
are harder (softer)   than  those 
with smaller (larger) $\psi$. 
It is known that 
this {\it elastic inhomogeneity} 
 gives  rise to   asymmetric 
elastic deformations in two-phase 
 structures and  eventual 
  pinning of 
the domain growth  \cite{Onukibook,Furukawa,Sagui}.

In our theory ${\Phi}(\psi, e_2, e_3 )$ in (2.7)  is  
 the simplest  periodic function of $e_2$ and $e_3$ 
with period 1. The periodicity arises from the fact that 
 the square lattice is invariant with 
respect to a slip of the crystal structure 
by a unit lattice constant 
along a line parallel to the $x$ or $y$ axis. 
Notice that, under rotation of the reference
frame by $\theta$, $e_2$ and $e_3$ 
are changed to  $e_2'$ and $e_3'$, respectively, 
with \cite{OnukiPRE}    
\bea
  e_2' &=& e_2 \cos 2 \theta + e_3 \sin 2 \theta, 
\nonumber\\
  e_3' &=& - e_2 \sin 2 \theta +  e_3 \cos 2 \theta .
\ena
For $\theta=\pi/2$ we have 
 $ e_2' = - e_2$ and $e_3' = - e_3$, so 
 $f_{\rm el}$ in (2.7)   remains invariant. 
For $\theta=\pi/4$ we have 
$ e_2' =  e_3$ and $e_3' = - e_2$ and recognize that 
the roles of tetragonal and shear strains are exchanged. 
For $\mu_2=\mu_3$,  the linear elasticity 
in (2.8) becomes isotropic, but  the nonlinear 
 elasticity is still anisotropic 
(from the fourth-order terms 
in the expansion of $\Phi$ in (2.7) 
 in powers of $e_2$ and $e_3$).

The elastic stress tensor 
$\asigma=\{\sigma_{ij}\}$  is 
 expressed as   
\bea 
\sigma_{xx} &=& K e_1+ \alpha\psi+ {\mu_2}
\sin(2\pi e_2)/2\pi ,\nonumber\\
\sigma_{yy}&=& Ke_1 +\alpha\psi-  {\mu_2}
\sin(2\pi e_2)/2\pi ,\nonumber\\ 
\sigma_{xy}&=& \sigma_{yx}= \mu_3
\sin(2\pi e_3)/2\pi . 
\ena  
In the linear elasticity,  
$\sin(2\pi e_2)/2\pi$ and 
$\sin(2\pi e_3)/2\pi$ are replaced by $e_2$ and $e_3$, respectively. 
 Notice the relation, 
\be 
\nabla\cdot\asigma= -
\frac{\delta}{\delta {\bi u}} F,   
\en 
where $\psi$ is fixed in the functional derivative 
$\delta F/\delta{\bi u}$.

The mechanical equilibrium condition  
$ 
\nabla\cdot\asigma= {\bi 0}
$  
is equivalent to the 
extremum condition $\delta F/\delta {\bi u}={\bi 0}$. 
In the coherent case this  condition  may be 
assumed  even in dynamics. 
In fact, using this condition in the linear elasticity,  
 the elastic field has been expressed in terms of $\psi$ 
in the previous theories (see the appendix)
\cite{Cahn71,Khabook,Fratzl,Onukibook,Cahn1,Cahn2}. 
We then find the following.
(i) The typical strain around domains   
is given by \cite{Onukibook}
\be 
e_0 = \alpha \Delta c/L_0, 
\en 
where $\Delta c= \Delta\psi/2$ is the composition difference 
between the two phases and 
\be 
L_0= K+ \mu_{20}
\en  
is the longitudinal elastic modulus. 
This strain needs to be small 
($e_0 <1/4$ approximately) as long as the system stays 
in the coherent regime. 
(ii) As will be shown in the 
the appendix, in the limit of weak cubic elasticity 
and weak elastic inhomogeneity, 
one-phase states become linearly 
unstable for 
$k_{\rm B}[T-T_0+ T\av{\psi}^2/2]/v_0  < \alpha^2/L_0$. 
At the critical composition $\av{\psi}=0$ 
this condition becomes  $T<T_{\rm s}$ with 
\be 
T_{\rm s}= T_0+ v_0\alpha^2/L_0k_{\rm B} . 
\en 
(iii) Furthermore, (A.5) suggests  that the typical domain size 
in steady  pinned states is 
a decreasing function of the 
quench depth $T_{\rm s}-T$.

\section{Dynamic equations}
\setcounter{equation}{0}

In the incoherent case 
the mechanical equilibrium does not hold  
 around dislocation cores
when dislocations are created 
and when they are moving \cite{OnukiPRE}. 
We thus need to set up 
the dynamic equation for the elastic displacement $\bi u$.  
In this paper  the lattice velocity 
${\bi v}= \p {\bi u}/\p t$ 
obeys the momentum equation \cite{Landau}, 
\be
 \rho \frac{\partial \bi{v}}{\partial t}
 = \eta_0\nabla^2 {\bi v}+ \nabla\cdot\asigma.  
\en 
The mass density $\rho$ and the shear viscosity $\eta_0$ 
are   constants. We neglect  the bulk viscosity  term 
($\propto \nabla\cdot{\bi v}$) in (3.1) 
for simplicity \cite{OnukiPRE}. 
In our model the sound waves relax 
owing to this viscous damping and the 
mechanical equilibrium 
$\nabla\cdot\asigma= {\bi 0}$ is rapidly attained 
unless $\eta_0$ is very small. Note that the nonlinear terms 
in (3.1) are only  those in $\sigma_{ij}$ in (2.12).

The composition obeys the diffusive equation, 
\be 
\frac{\partial \psi}{\partial t} = 
\nabla \cdot  \lambda (\psi) \nabla
  \frac{\delta {F}}{\delta \psi}. 
\en
The kinetic coefficient depends on $\psi$ 
as \cite{Kitahara,Binder} 
\be 
\lambda(\psi)= 
\lambda_0( 1 - \psi^2 )= 
4\lambda_0c_Ac_B, 
\en 
where $\lambda_0$ is a constant. 
Here $\bi u$ is fixed in the chemical potential difference 
$\delta F/\delta \psi$, so 
\bea 
\frac{\delta {F}}{\delta \psi} &=& 
\frac{k_{\rm B}}{v_0} \bigg [ \frac{T}{2} \ln \bigg (\frac{1+\psi}{1-\psi} \bigg ) - 
T_{0} \psi\bigg ]  -C \nabla^2\psi +\alpha e_1  \nonumber\\
&+& 
\frac{\mu_{21}}{4\pi^2}[ 1 - \cos (2 \pi e_2)] 
+\frac{\mu_{31}}{4\pi^2} [ 1 - \cos (2 \pi e_3)] .  
\ena 
The last two terms arise from the elastic inhomogeneity. 
If $\lambda (\psi)$ is of the form of (3.3),  
the diffusion equation 
$\p c_A /\p t= D_0 \nabla^2 c_A$ ($\p c_B /\p t= D_0 \nabla^2 c_B$) follows 
in the dilute limit $c_A\rightarrow 0$ 
($c_B\rightarrow 0$) with 
\be 
 D_0=\lambda_0 k_{\rm B}Tv_0^{-1},
\en 
where the coupling to the elastic field becomes negligible.  
In usual solid mixtures  
 the  diffusion is very slow 
and  vacancies are in many cases 
crucial for a  microscopic description 
of diffusion  \cite{Binder}. 
Effects of such point defects are not treated  in 
the present theory.

The  total free energy $F_{\rm tot}= 
F+ \int d{\bi r} \rho {\bi v}^2/2$ including the kinetic energy 
then changes in time as 
\be 
\frac{d}{dt}F_{\rm tot} = 
- \int d{\bi r} \bigg [ \sum_{ij}\eta_0 (\nabla_i v_j)^2 
+ \lambda (\psi)  \bigg | \nabla \frac{\delta {F}}{\delta \psi}\bigg |^2
\bigg ].
\en   
Here  the surface integrals have  been omitted, which 
vanish if the boundaries are fixed and there is no flux of 
the atoms from outside 
(and also if the periodic boundary condition 
is imposed in simulations).  
The above time-derivative is 
non-positive-definite. As a result,
the equilibrium is attained when 
${\bi v}= \nabla\cdot\asigma
={\bi 0}$ and $\delta F/\delta \psi=$const.

If the lattice is deformed  significantly, 
we should add the convective term 
 $-\nabla(\psi{\bi v})$ on the right hand side
 of  (3.2), treating (3.2) as the equation in the 
Euler description. If its presence 
 is assumed,  another term of the form,   
$-\psi\nabla \delta F/\delta\psi$, 
becomes also needed on 
the right hand side of (3.1). 
With these two terms we again 
have $dF_{\rm tot}/dt \le 0$. 
These two terms are 
well-known in critical dynamics of fluids 
\cite{Onukibook}. However, in our solid case, 
the magnitude of the 
displacement $\Delta{\bi u}= \int_0^t  dt' {\bi v}
({\bi r},t')$ remains   small 
and these two terms  give rise 
to no essential differences  
in our results at not large applied strains 
\cite{convection}.

Also note that the dynamic equations (3.2) and (3.3) may be treated 
as Langevin equations with addition of the 
random noise terms related to the kinetic coefficients 
$\lambda (\psi)$ and $\eta_0$ via the fluctuation-dissipation 
relations \cite{Onukibook}.
In this paper, however, we  neglect the random noise, 
because  the thermal energy $k_{\rm B}T$ 
will be   assumed to be much smaller than 
the typical energy of elastic deformations.

\section{Numerical Results} 
\setcounter{equation}{0}

\subsection{Method} 

We integrated (3.1) and (3.2) in two dimensions 
on a $256\times 256$ square lattice. 
The mesh size $\Delta x$ was set equal to 
the lattice constant $a$ 
in the reference state with ${\bi u}={\bi 0}$, 
so the system length is $L_0=256a$.  
The vectors, $\bi u$ and $\bi v$,  
are defined at the lattice points  $(n,m)$, while 
the strains, the tensors, and the composition 
are defined  on the middle points 
$(n+1/2,m+1/2)$. These are   needed  to realize   
well-defined microscopic slips 
in our numerical scheme \cite{OnukiPRE}. 
The periodic boundary condition 
was imposed except  the simulation 
of applying uniaxial deformation 
(Figs.13 and 14).  
Because the time scale of $\bi u$ is shorter 
than that of $\psi$, 
 we integrated  (3.1) using 
an implicit Crank-Nicolson  method. 
Space and time will be measured in  units 
of $a$ and  
\begin{equation}
\tau_0 = (\rho/\mu_{20})^{1/2}a,
\end{equation}
respectively, where $\mu_{20}$ is defined by (2.10) and 
$(\mu_{20}/\rho)^{1/2}$ is the  transverse sound velocity 
propagating in the $[11]$ direction. 
The  free energies 
 and the free energy  densities 
 are measured in units of 
 $\mu_{20}a^2$ and $\mu_{20}$, respectively. 
For simplicity, 
the scaled time $\tau_0^{-1}t $, 
position vector $a^{-1}{\bi r}$, 
and displacement vector  $a^{-1}{\bi u}$ 
will  be written as  $t$, ${\bi r}$, and ${\bi u}$, 
respectively, 
in the same notation.

In this paper we set 
$K/\mu_{20}=4.5$, $\alpha/\mu_{20}=0.6$,  and 
$k_{\rm B}T_0/v_0 \mu_{20}= 0.05$, where 
$T_0$ is the mean-field critical temperature in 
(2.4). Since  $T \sim T_0$ 
hereafter,   the elastic energy to create a single 
slip  ($\sim \mu_{20}v_0$)  is much larger than $k_{\rm B}T$ 
 in our simulations. 
Furthermore,  we assume  weak  
cubic elastic anisotropy with  
$\mu_{30}/\mu_{20}=1.1$ 
and  moderate  elastic inhomogeneity with  
$\mu_{21}=\mu_{31}=0.6\mu_{20}$.

The dimensionless kinetic coefficients are  given by 
\be 
\lambda_0^*= \lambda_0\tau_0\mu_{20}a^{-2}, \quad 
\eta_0^*= \eta_0 /\tau_0 \mu_{20}.
\en 
We set 
 $\lambda_0^*=0.001$  and   $\eta_0^*=0.1$. 
Then, 
\be 
\lambda_0^*/\mu_0^*= D_0\rho /\eta_0 \sim 10^{-2}. 
\en  
Since the relaxation rate of a sound
 with wave number $k$ is  
$\eta_0 k^2/\rho$, the time scale of $\psi$ becomes  
 longer than that of  the elastic field by 
two orders of magnitude.  In 
real solid alloys, these two time scales are 
much more distinctly separated, 
probably  except for 
hydrogen-metal systems 
where the protons diffuse quickly 
\cite{Onukibook}.

In homogeneous one-phase states 
we have $e_2=e_3=0$ and $e_1= -\alpha\psi/K$. 
Here  well-known is a parameter $\eta= |\p a/\p \psi|/a$ 
representing   the strength of the 
composition-dependence  of 
the lattice constant $a$ in a mixture \cite{Cahn1}. 
In our case we have 
 $\eta= \alpha/2K= 0.067$ 
and the spinodal temperature 
$T_{\rm s}$ in (2.16) becomes 2.31.

\subsection{Slips and composition changes}

Edge dislocations appear in the form of slips or 
dipole pairs \cite{OnukiPRE}, 
because a single isolated 
dislocation requires a very large elastic energy. 
Slips are  thus fundamental units of plastic deformations. 
In Fig.2  we show the displacement 
and the composition around typical slips 
in a one-phase steady state with length 
$10\sqrt{2}a$ in the upper plate and $10a$ 
in the lower plate. Here  we initially prepared  
a slip given by the linear elasticity 
theory \cite{OnukiPRE}  at the critical composition 
($\av{\psi}=0$) and let $\bi u$ and $\psi$ 
relax until the steady state was achieved. 
The temperature was kept at $T/T_0=2.5$ 
and no phase separation occurred.   
As in the previous simulations \cite{Chen,Sro}, 
we can see Cottrell atmospheres around 
the dislocation cores. The  
 maximum and  minimum 
of $\psi$ at the lattice points close to the 
dislocation cores are of order 
$\pm  0.6$.  Cottrell's result is obtained as follows: 
Let  $T$ be much higher than $T_0$ and $\alpha^2/L_0$  
and the gradient term 
be  neglected; then,  
the condition $\delta  F/\delta\psi=$const. 
yields $c_A/(1-c_A) =$const.
$\exp (- U/k_{\rm B}T)$, where $U=v_0\alpha e_1$ 
\cite{comment2}. 
In our case the maximum of $|U|/k_{\rm B}T$ 
at the lattice points is of order 1 and the accumulation is not 
very strong.

As a next step, 
starting with  the configuration in Fig.2, 
we  lowered the temperature 
to  $T/T_0= 2$ to induce spinodal decomposition. 
Subsequently the Cottrell atmospheres  
grew into domains 
and  the  dislocation cores stayed 
 at the interface regions. 
The  domain size attained finally 
was   of order $50a$.   Fig.3 illustrates 
 the displacement and the composition 
in the final steady state, where the maximum and minimum of $\psi$ are 
about $\pm 0.9$. 
L{\'e}onard and Desai \cite{Desai} obtained 
similar composition profiles 
in spinodal decomposition, where  the elastic field 
of dislocations 
(given by the linear elasticity theory) 
was fixed in space and time.

Mathematically, slips in steady states  satisfy   
$\delta F/\delta {\bi u}={\bi 0}$ 
and $\delta F/\delta \psi=$const. 
Without externally applied strains, 
they are metastable owing  to the Peierls potential  energy 
arising from the discreteness of 
the lattice structure \cite{OnukiPRE}. 
Although not discussed in this paper, 
slips become unstable against   expansion   or 
shrinkage  with increasing  applied strain.

\subsection{Dislocation formation around a  hard domain}

Fig.4 shows  a  single large hard (A-rich) domain at the center 
in the coherent condition 
at shallow quenching $T/T_0=2$ 
after a long equilibration time. 
Here  $\psi$ is  about 
$0.7$ inside the domain 
and  about $-0.7$ outside it. Its shape slightly deviates from 
sphericity owing to the weak cubic anisotropy assumed 
in this paper.  We next 
performed a second deeper quenching  to  
$T/T_0= 1$.  Subsequent diffusional 
adjustment of the composition proceeded   very slowly, 
but a discontinuity of the order parameter $\Delta \psi$ 
about 1.8 was established 
relatively rapidly across  the interface \cite{Onukibook}. 
As a result, at  a time about 1000 
after the second quenching,  
the maximum of $|e_2|$ reached $1/4$, 
 the value at the stability limit, 
in the interface region (see the sentences below (2.7)).
We then observed formation of dislocations  
and generation of sound waves emitted from the dislocations.  
The upper panel of Fig.5 shows  the coherent 
elastic displacement ${\bi u}_{\rm coh}$ 
just before the dislocation formation, 
while the lower panel shows the subsequent additional 
incoherent change 
$\delta{\bi u}= {\bi u}-{\bi u}_{\rm coh}$ after a  
 time interval  of 1000. The free energy $F$ in the state in 
the lower panel is smaller than that in the upper panel 
by 152.9 in units of $\mu_{20}a^2$. 
More details are   as folllows: 
(i) two pairs of dislocation dipoles (four dislocations) 
appeared simultaneously  in a narrow region,  
(ii)  two of them 
glided  preferentially  into the 
softer region forming two slips perpendicular to each other, 
and (iii) slips collided in many cases 
and  stopped  far from the droplet, resulting in 
a nearly steady elastic deformation. 
Thus a half of the dislocation 
cores stayed  at the interface and 
the others were distributed around  the domain. 
These three processes took only  a short 
time of order 100.

After the above dislocation formation 
at a relatively early stage, 
the composition changed very slowly.
We show three figures at  $t=23000$.
In Fig.6 we displays  the following strain, 
\be 
e= ({e_2^2+e_3^2})^{1/2},   
\en 
which is invariant with respect to 
the rotational transformation (2.11). 
The slips make an angle of $\pm \pi/4$ 
with respect to the $x$ axis 
in the regions with large $|e_2|$ (in the  uniaxially 
deformed regions), 
while they are parallel 
to the $x$ or $y$ axis  in the 
corner regions with large 
$|e_3|$ \cite{OnukiPRE}. 
We also notice that  
the dislocation formation took place 
with the symmetry axis 
in the $[11]$ direction for our special  geometry. 
Fig.7 gives  the free energy density 
$f$ in (2.3), where the peaks outside the domain 
represent  the dislocation cores.  
In  the interface region it  exhibits 
a cliff-like structure   arising from the 
gradient term  and  higher peaks  arising from 
 the dislocation cores.   
Fig.8 shows the order parameter $\psi$, 
where we can see 
Cottrell atmospheres around the dislocation cores 
surrounding the domain. The system is 
still transient 
and there is still a small composition 
flux through the interface.

\subsection{Dislocation formation in  a soft network}

Next we examine dislocation formation 
when hard rectangular domains are 
densely distributed and wrapped by a percolated 
soft network. As in Fig.9, we prepared such a  steady domain 
structure at $T/T_0=2$ in the coherent condition. 
As in the previous simulations \cite{Furukawa,Sagui},  
the hard domains (in gray) are elastically 
isotropic, while  the soft network (in white) 
is mostly uniaxially stretched. 
That is, in the soft stripes  between 
the two adjacent hard domains,  we obtain 
$e_2 \sim 0.2$ in the horizontal stripes 
and  $e_2 \sim -0.2$ in the vertical stripes. 
We then quenched $T$ to $T/T_0= 1$ to 
induce the composition readjustment. 
 Fig.10 displays  the resultant time evolution of 
the total free energy  $F=\int d{\bi r}f$ and the snapshots of 
$e$ in (4.4) at the points 
A, $\cdots,$ and  E. It  demonstrates 
that $F$   mainly decreases due to the composition change 
but sometimes due  to appearance and gliding of slips in the soft 
 stripes.  Note that the overall composition adjustment 
occurs  slowly on the time scale of $R^2/D_0=10^5-10^6$ 
where $R$ is the domain size. 
In Fig.11 we show the displacement $\bi u$
 within  the square window in B, C and D, respectively,  
while in Fig.12  the bird views of 
the free energy density $f$ the square window  are given 
at $t=0$ and 4475 after the second quench. 
 Fig.12    clearly illustrates   appearance 
of the  peaks representing  the dislocation cores.

\subsection{Uniaxial stretching  in two-phase states}

Finally we apply a constant 
uniaxial deformation to initially coherent 
states  with $\av{\psi}=0$ 
to induce plastic flow. 
That is,   we set $u_x=u_y=0$ at the bottom ($y=0$) 
and  $u_x=-u_y=\epsilon L_0/2$ at the top ($y=L_0$). 
The applied strain rate was fixed at 
$\dot{\epsilon}=10^{-4}$, so 
$\epsilon=  \dot{\epsilon} t$ 
with  $t$ being the time after application of the deformation. 
In  Fig.13 we plot the average 
normal stress $N_1$ vs the applied strain $\epsilon$ 
for $T/T_0= 3$, $2.4$, and $2$ (upper plate), where   
\be 
N_1= 
\av{\sigma_{xx}-\sigma_{yy}}
= \frac{1}{\pi}
\av{\mu_2 \sin (2\pi e_2)},
\en 
where $\av{\cdots}$ denotes taking the spatial average. 
The snapshots of $e$ in (4.4) are also given 
at the points a, b, and c (lower plates). 
For $T/T_0 =3$ the system is  
in a homogeneous one-phase state  
and  random numbers  with variance 
0.01 were  assigned to $\psi$ at the lattice points at $t=0$.  
In the initial state at 
$T/T_0 =2.4$ the maximum and minimum of $\psi$ and $e_2$ 
are $\pm 0.32$ and $\pm 0.05$, respectively.  
At  $T/T_0 =2$ these numbers are magnified to 
$\pm 0.75$   and 
$\pm 0.20$. All the initial states are coherent without 
dislocations.  For  $T/T_0 =3$ the elastic 
instability occurs at $\epsilon=1/4$ resulting in a fine 
mesh of slips as in the lower left plate. 
For  $T/T_0 =2.4$   the onset point of the slip formation 
is decreased to  $\epsilon=0.17$. 
For  $T/T_0 =3$  the onset is very early at 0.015, 
the stress-strain  relation
 exhibits  zig-zag behavior upon  appearance of slips, 
and the stress continue to increase on the average 
(up to the upper bound of $\epsilon$ given by  0.35 in the simulation). 
 Fig.14 consists of  snapshots of  $e$ in (4.4) 
and the shear deformation  energy density $\Phi$ 
in (2.7) in units of $\mu_{20}$. 
We can see quadratic appearance 
of dislocations 
at the center of the uniaxially stretched  stripes 
at $\epsilon=0.05$ (top plate), gliding of the dislocations 
and pinning at the interfaces at $\epsilon=0.1$ 
(top and middle plates), 
and thickening of the slips into {\it shear bands} 
at $\epsilon=0.2$ (bottom plate).

We mention a creep 
 experiment in the presence of 
high volume fractions of $\gamma'$ 
precipitates  \cite{pollock},  
 where softer disordered $\gamma$  
regions  were  observed to be 
filled with dislocation networks 
after large deformations.

\section{Summary and Concluding Remarks}
\setcounter{equation}{0}

In summary, we have presented a coarse-grained 
phase field model of plastic deformations in two-phase 
alloys.
Though our simulations have been  
performed in two dimensions, 
 a number of insights into the very complex 
processes of plasticity  have been gained. 
We mention them and give some remarks. \\
(i) Performing a  two-step 
quench, we have numerically examined  dislocation 
formation around the interface regions, which occur 
spontaneously  in deeply quenched 
phase separation. Experimentally \cite{loss}, 
dislocation formation has been observed 
around  growing 
$\gamma'$(Al$_3$Sc) precipitates  
at low volume fractions 
when  the  radii exceeded  a threshold 
about 20nm.  Such  spontaneous dislocation 
formation  with domain growth has not yet been 
studied theoretically. 
\\ 
(ii) We have found that  dislocations glide 
preferentially into the softer regions 
with smaller shear moduli 
and tend to be trapped 
in the interface regions in agreement 
with a number of observations \cite{Strudel}. 
Theoretically, the composition-dependence 
of the elastic moduli (elastic inhomogeneity) 
is a crucial ingredient  to explain the experiments.  
\\
(iii) We have  applied uniaxial strain to 
create multiple slips in 
two-phase alloys which were  initially in 
the coherent condition. 
The dislocation formation starts in the 
mostly stretched middle points 
of the soft stripes. 
A stress-strain curve 
in Fig.13 at deep quenching   
is very different from the curves 
in one-phase states. 
In  real two-phase alloys, a 
  similar  monotonic  increase  
of the stress without overshoot 
has  been observed, but  a considerable  amount of 
 defects  should  preexist 
in such experiments particularly in work-hardened samples  
\cite{Strudel,Cottrell,Haasen}. 
\\

This work is  a first theoretical step to understand  
 complex phenomena of incoherency in solids. 
Finally, we mention two future problems 
which could  be  studied numerically 
in our scheme.\\
(i) The composition has been  taken as a single 
order parameter. 
Extension of our theory is needed 
to more general phase separation processes 
 involving an 
order-disorder phase transition \cite{Khabook,Onukibook,Desai}
 and to diffusionless 
(Martensitic) structural phase 
transitions \cite{Khabook,Onukibook}.\\
(ii) Dislocations move under applied strain. 
The motion is  complicated when 
they are coupled with an order parameter 
and when the time scale of the order parameter 
is slow  \cite{Sro,Bausch}.

\vspace{2mm} 
{\bf Acknowledgments}
\vspace{2mm}

We   would like to thank  
Toshiyuki Koyama  for valuable discussions 
 on the incoherency effects in metallic alloys.  
This work is supported by 
Grants in Aid for Scientific 
Research 
and for the 21st Century COE project 
(Center for Diversity and Universality in Physics)
 from the Ministry of Education, 
Culture, Sports, Science and Technology of Japan.

\vspace{2mm} 
{\bf Appendix}\\
\setcounter{equation}{0}
\renewcommand{\theequation}{A.\arabic{equation}}

Here we assume weak 
elastic anisotropy and weak elastic 
 inhomogeneity in the coherent condition in two dimensions, 
supposing shallow quenching.  
Then we may eliminate the elastic field 
in terms of $\psi$  using the mechanical equilibrium condition 
$\nabla\cdot{\tensor{\sigma}}={\bi 0}$ in the linear elasticity. 
We consider the space integral of the 
last two terms in  the free energy density  in (2.3):  
$\Delta F =  \int d{\bi r}
 [ \alpha e_1\psi +   f_{\rm{el}}  ]$. 
We assume that  $|\mu_{21}|$ and $|\mu_{31}|$ 
are much smaller than $L_0= K+\mu_{20}$ 
and that  $\xi_a=  2(\mu_{20}/\mu_{30}-1)$ is  small. 
Then $\Delta F$  may be rewritten as  
 \cite{Onukibook,Furukawa,Sagui}  
\bea 
\Delta F 
&=&   \int d{\bi r} \bigg [- \frac{\alpha^2}{2L_{0}}\psi^2 
+\frac{1}{2} \tau_{\rm cub} |\nabla_x\nabla_y w|^2 \bigg ] \nonumber\\
&+&  
 \int d{\bi r} \bigg [ g_{2} 
\psi  |(\nabla_x^2-\nabla_y^2)w|^2+ 
 g_{3} \psi | \nabla_x\nabla_y w |^2  \bigg ],  
\ena
where    $w$ is obtained from 
the Laplace equation, 
\be 
\nabla^2w= \psi-\av{\psi},
\en 
with $\av{\psi}$ being the  average order parameter. 
In the first line of (A.1)  the bilinear terms are written with  
\be 
\tau_{\rm cub}= -({2\alpha^2}/{L_0^2}) \mu_{20} \xi_a.  
\en
The term proportional to 
$\tau_{\rm cub}$ gives rise to anisotropic domains \cite{Nishimori}. 
The second line consists of  the third-order terms with 
\be  
g_{2} = \mu_{21}\alpha^2/2L_0^2,\quad 
 g_{3} = 2\mu_{31}\alpha^2/L_0^2.  
\en 
The third-order  terms  are known to give rise to pinning of 
domain growth (and some frustration effects when 
$g_2$ and $g_3$ have different signs) \cite{Furukawa,Sagui}.

In our simulations 
we set $\xi_a= 2(1/1.1-1)\cong - 0.18$ 
 and 
$\tau_{\rm cub}\cong 0.0043\mu_{20}$, 
so the domains tend to 
become square or rectangular  
with interfaces parallel to the $x$ or $y$ axis. 
Furthermore, we set  $g_2=g_3/4\cong 
0.0035\mu_{20}$. 
For $\mu_{21}\sim \mu_{31}$ 
the typical domains in pinned two-phase states $R_{\rm E}$ 
 is given by \cite{Furukawa} 
\be 
R_{\rm E} 
 \sim \gamma /[\mu_{21}(\Delta c)^3],  
\en 
where $\gamma$ is the surface tension and 
 $\Delta c$ is the composition difference 
between the two phases. 
Thus $R_{\rm E}$ decreases as the quenching  becomes deeper.



\newpage 

\begin{figure}[t]
\includegraphics[scale=0.8]{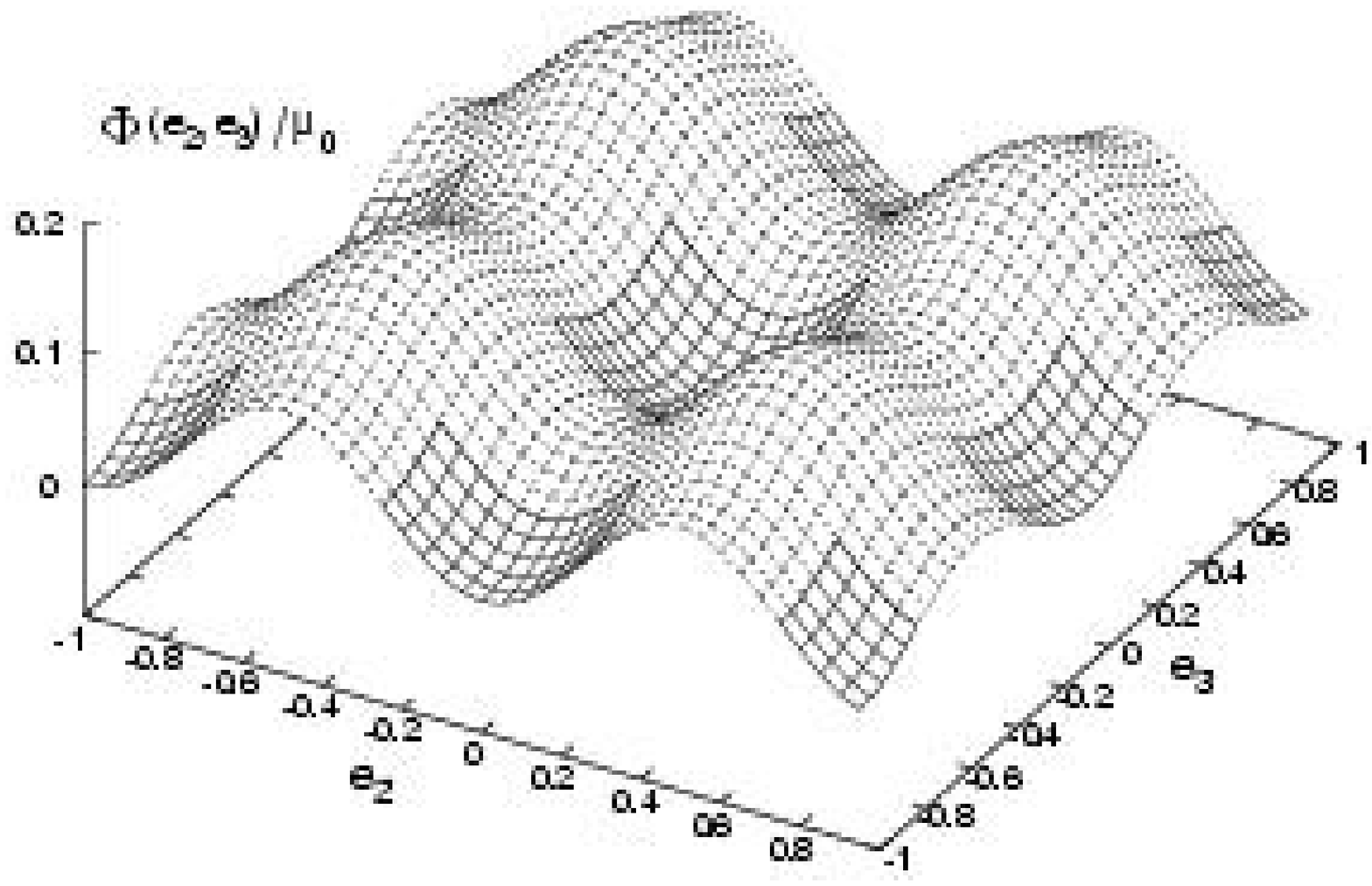}
\caption{
Normalized shear deformation energy 
$\Phi(e_2,e_3)/\mu_0$  for the case 
$\mu_2=\mu_3=\mu_0$. 
The elastically stable regions are  meshed 
 with solid lines on the surface.   
 }
\label{1}
\end{figure}

\begin{figure}[t]
\includegraphics[scale=0.8]{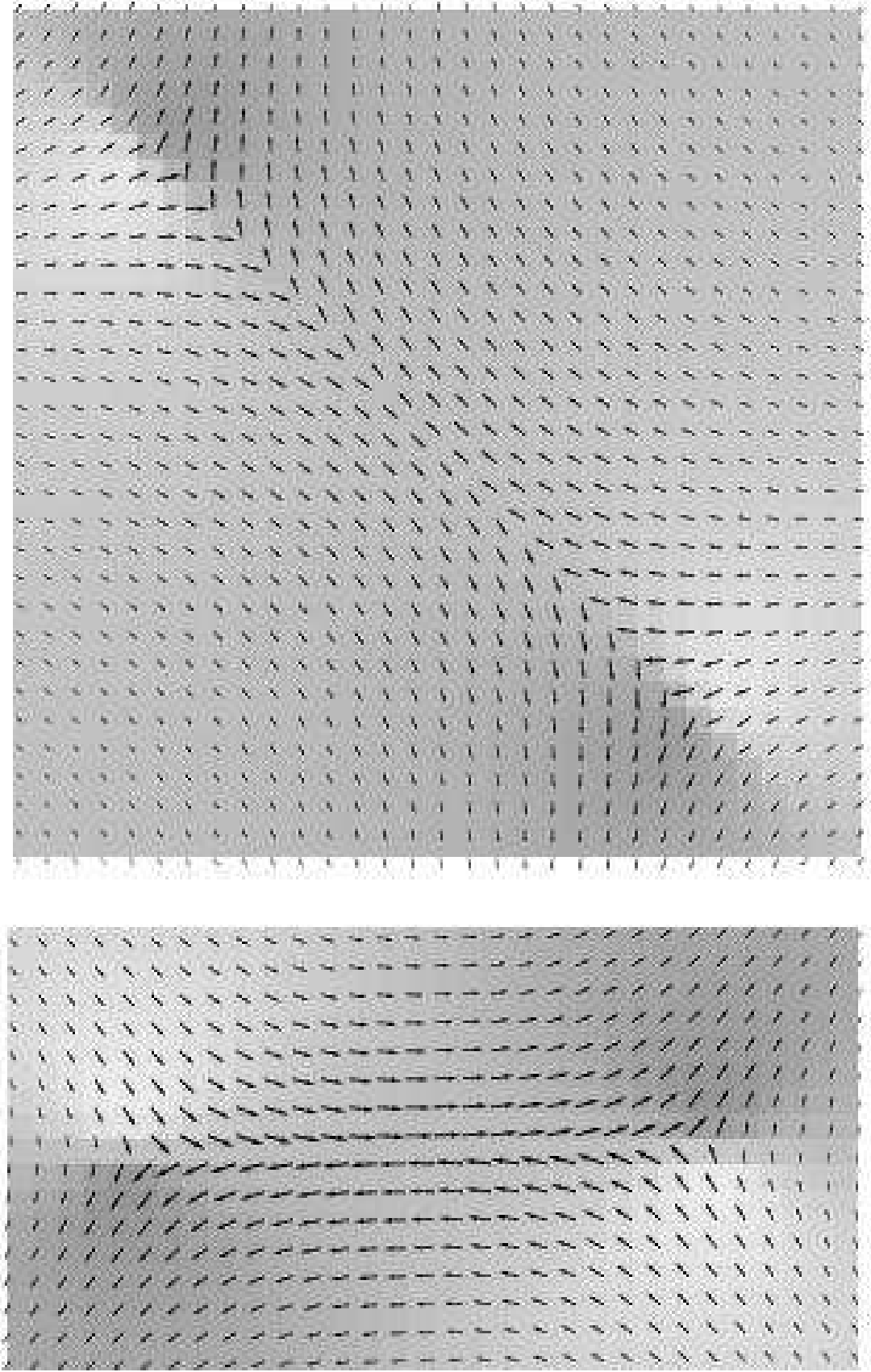}
\caption{
Displacement vector 
for a slip (dislocation pair) 
making an angle of $3\pi/4$ (upper plate) 
and $0$ (lower plate) 
with respect to the $x$ (horizontal) axis 
in a one-phase steady state at $T/T_0=2.5$.
The arrows are from the initial position 
in a perfect crystal to the deformed position. 
The degree of darkness represents 
the composition.
}
\label{2}
\end{figure}

\begin{figure}[t]
\includegraphics[scale=0.9]{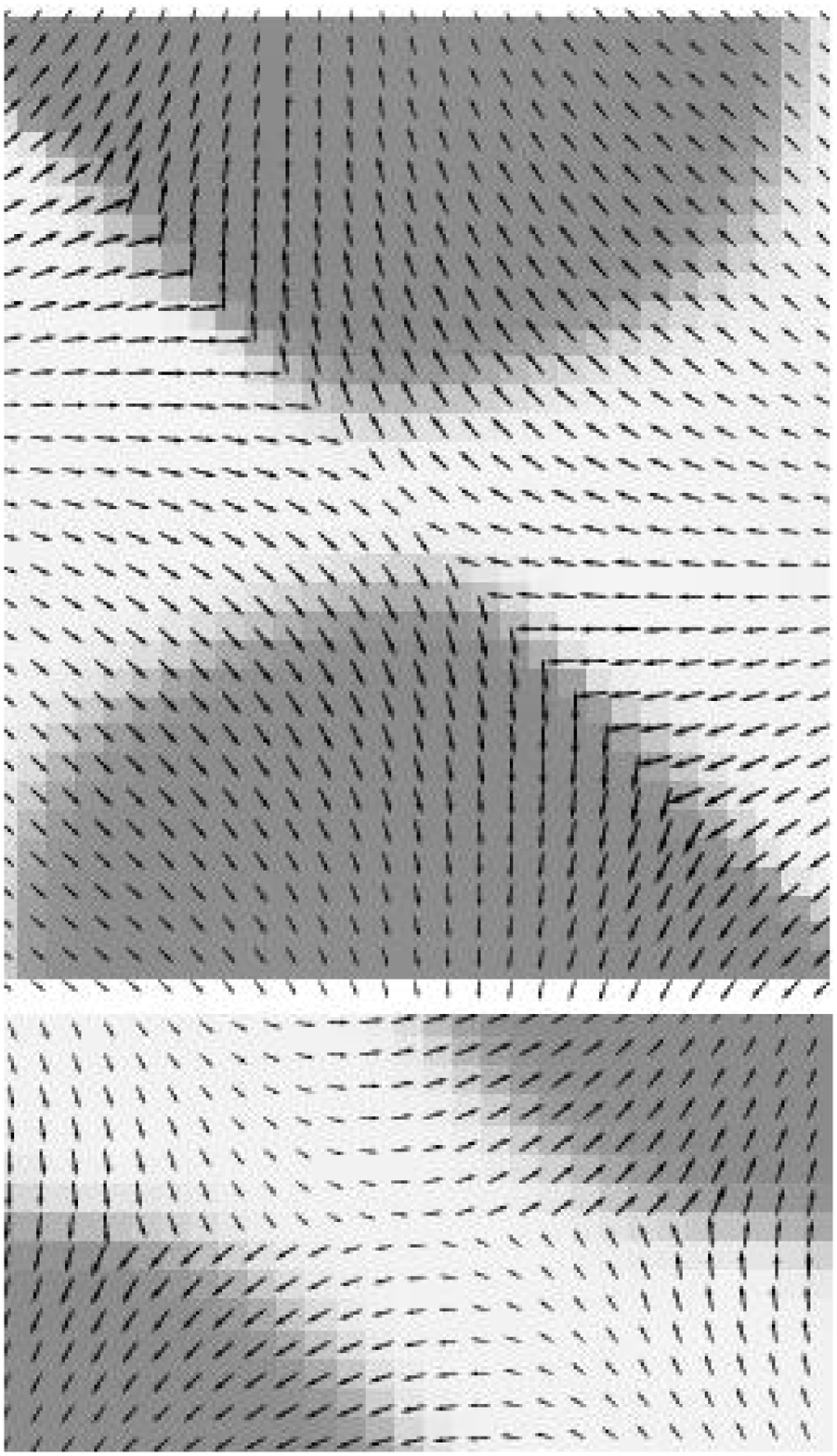}
\caption{
Displacement vector for a slip (dislocation pair) in 
 a two-phase steady state at $T/T_0=1.7$  
obtained from the configuration in Fig.2 
after quenching. The arrows are from the initial position 
in a perfect crystal to the deformed position. 
The dislocation cores 
 are trapped at the 
interface regions.}
\label{3}
\end{figure}

\begin{figure}[t]
\includegraphics[scale=0.8]{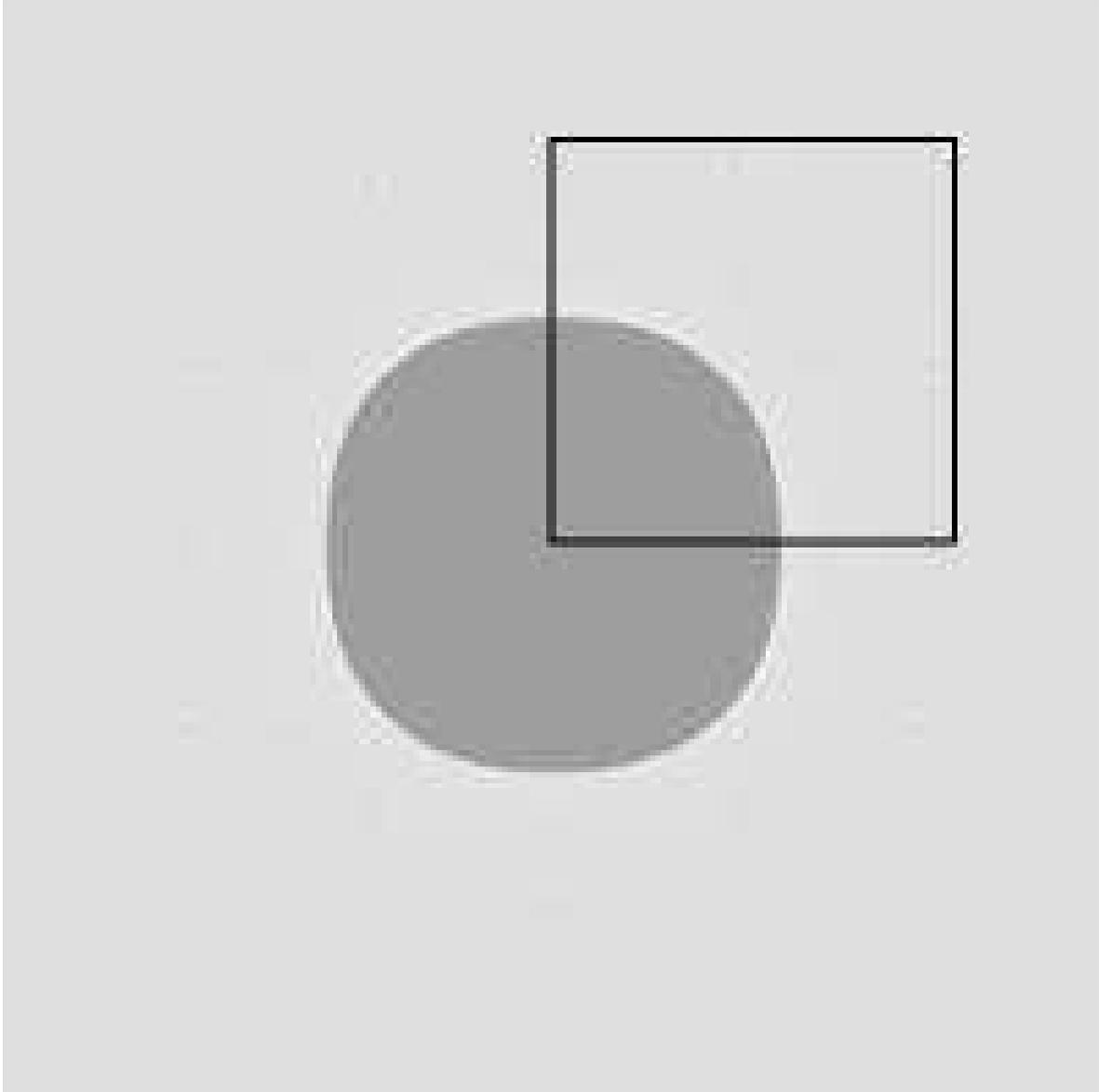}
\caption{
Hard domain in a steady state  at $T/T_0=2$. 
The displacement in the square  region 
will be  displayed in Fig.5.}
\label{4}
\end{figure}

\begin{figure}[t]
\includegraphics[scale=0.7]{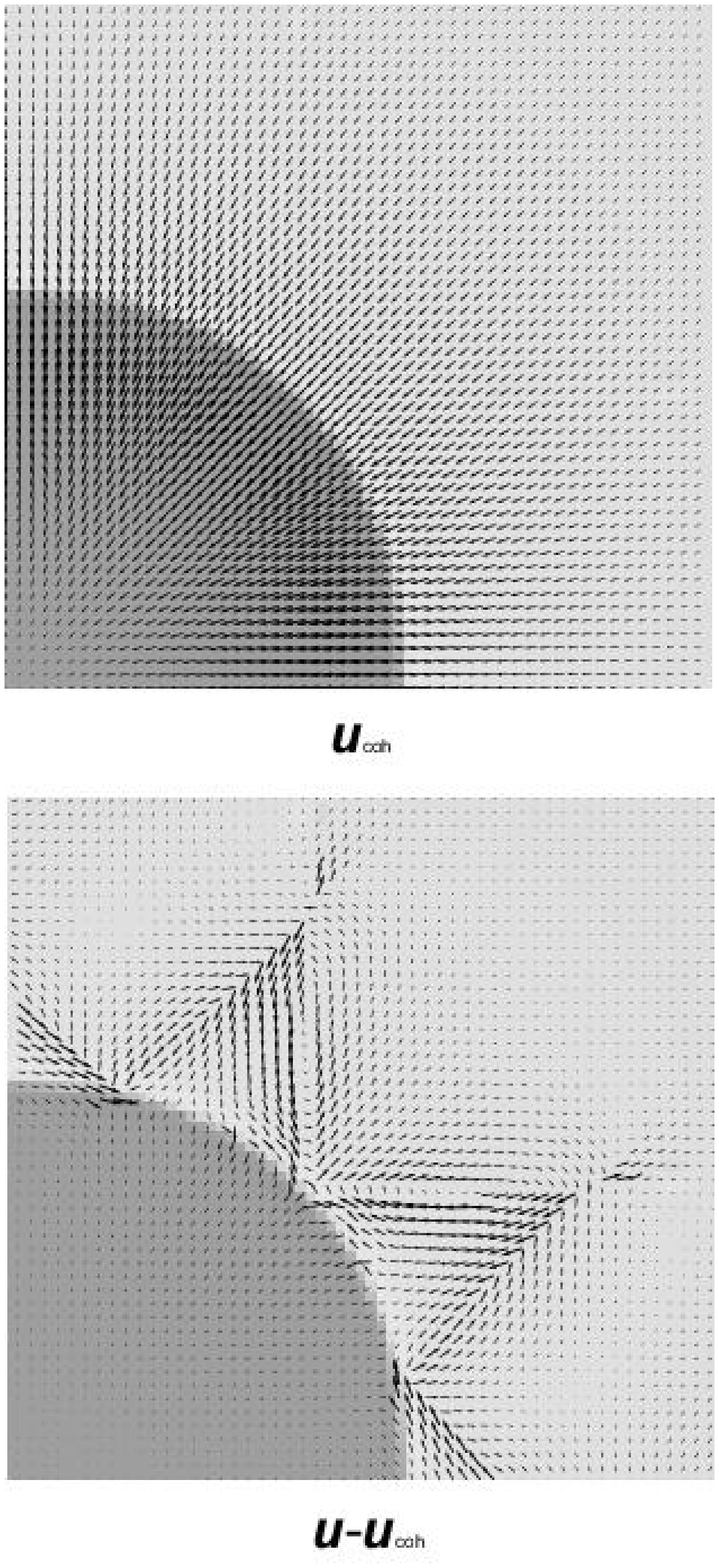}
\caption{
Upper plate: Coherent elastic displacement  
${\bi u}_{\rm coh}$ just 
 before birth of dislocations at  deep quenching 
at $T/T_0=1$. 
Right: Incoherent elastic displacement after 
appearance of dislocations. 
}
\label{5}
\end{figure}

\begin{figure}[t]
\includegraphics{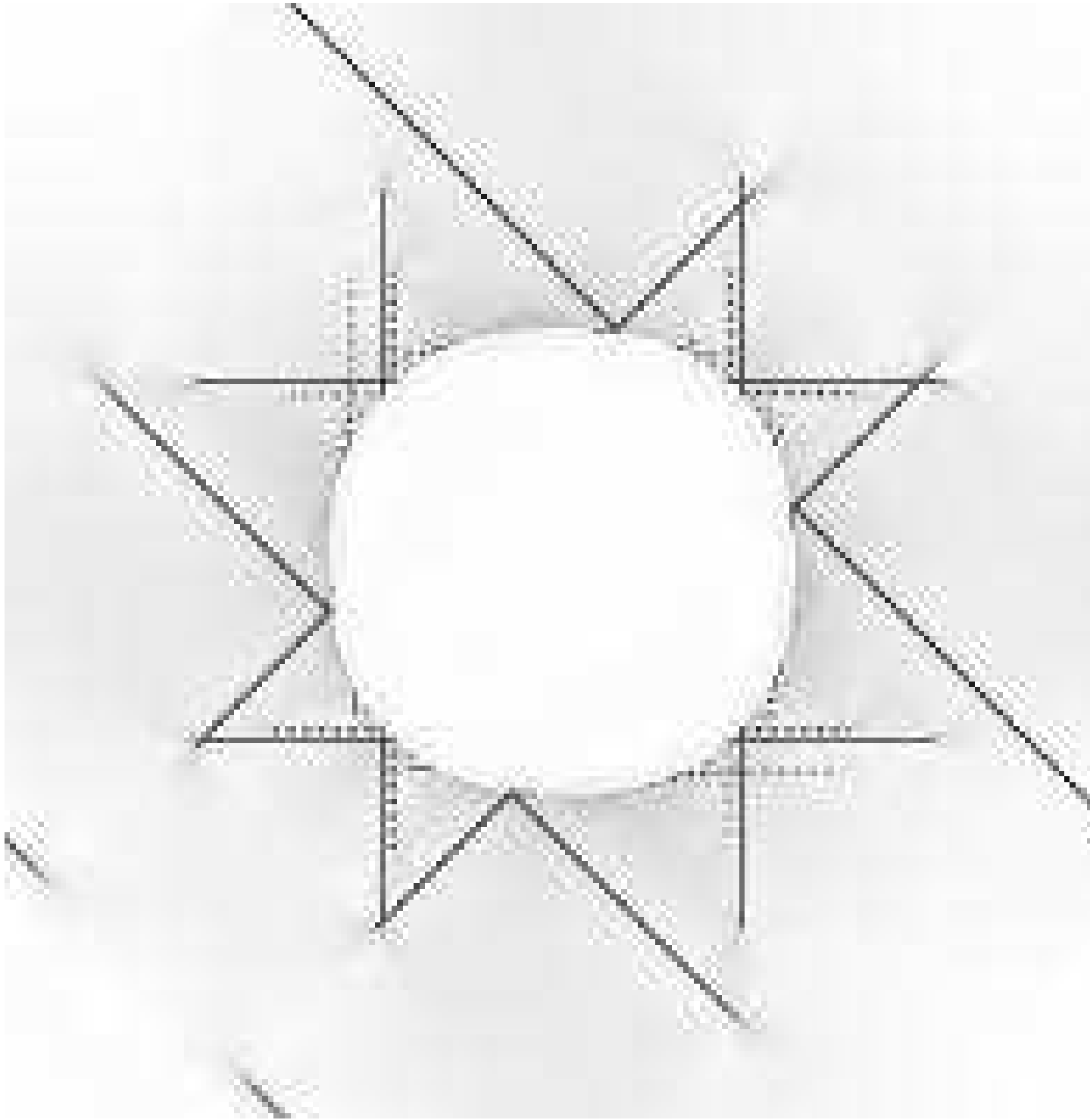}
\caption{
Snapshot of $e$ in (4.4) after dislocation formation, 
which is zero within  the hard domain and nonvanishing 
outside. The slip lines end at dislocations.    
}
\label{6}
\end{figure}

\begin{figure}[t]
\includegraphics[scale=0.8]{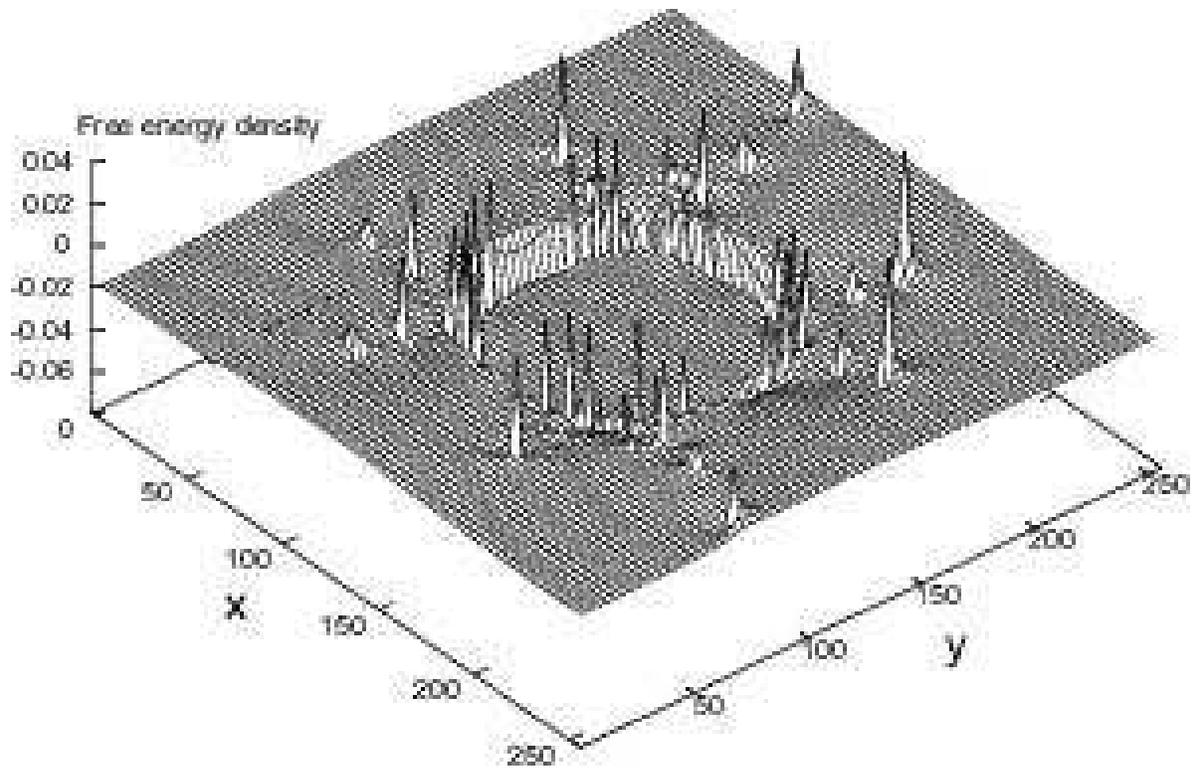}
\caption{
Snapshot of the free energy density $f$ 
in (2.3). The peaks are located near 
 the dislocation cores 
and the  cliff-like structure 
 represents the interface free energy density.    
 }
\label{7}
\end{figure}

\begin{figure}[t]
\includegraphics[scale=0.8]{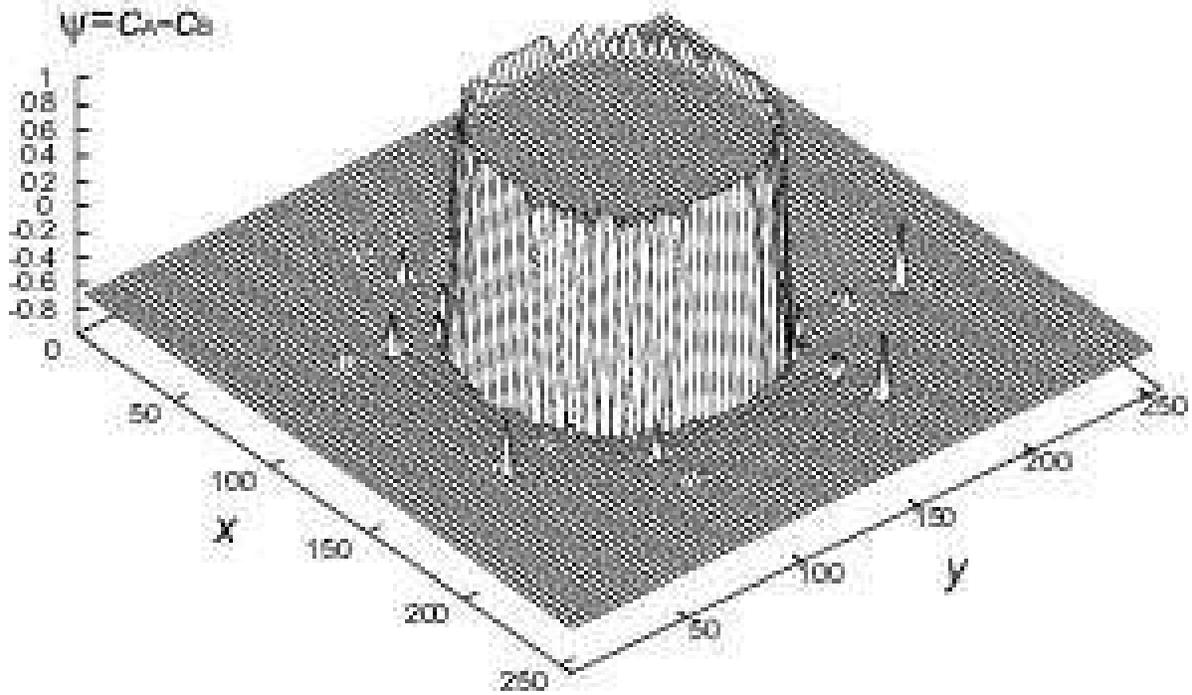}
\caption{
Snapshot of the order parameter 
around a  hard domain in the incoherent case 
obtained after  a two-step quench. 
The  peak  structure 
 at the interface arises because 
the system away  from the interface 
is still in a  transient state. 
The peaks around 
 the dislocation cores in the outer soft region 
  represent Cottrell atmospheres 
(but the minima paired  are not seen in the figure). 
 }
\label{8}
\end{figure}

\begin{figure}[t]
\includegraphics{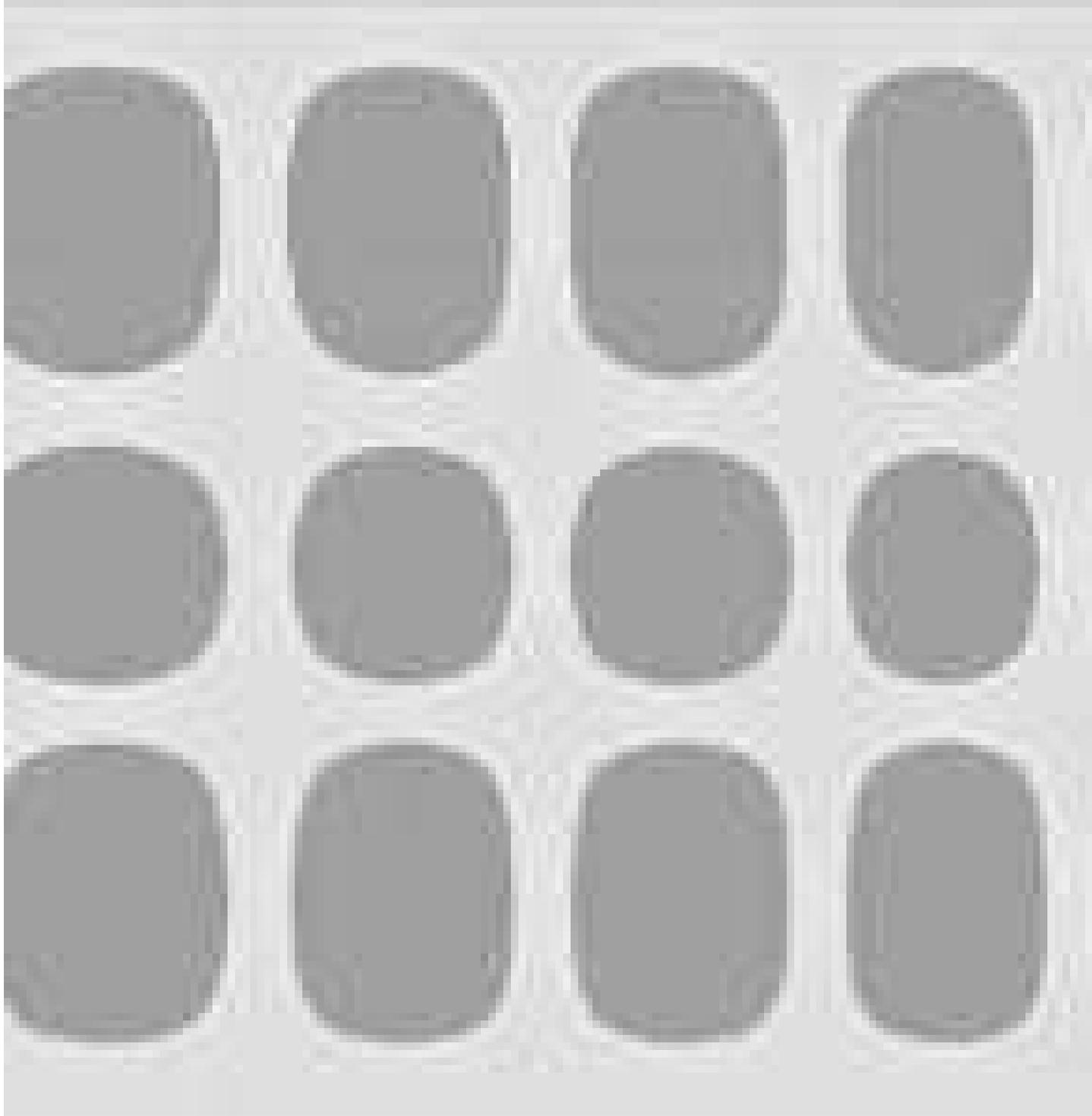}
\caption{
Domain structure obtained 
at a  shall quench $T/T_0= 2$ 
in the coherent condition. 
 }
\label{9}
\end{figure}

\begin{figure*}[t]
\includegraphics[scale=0.9]{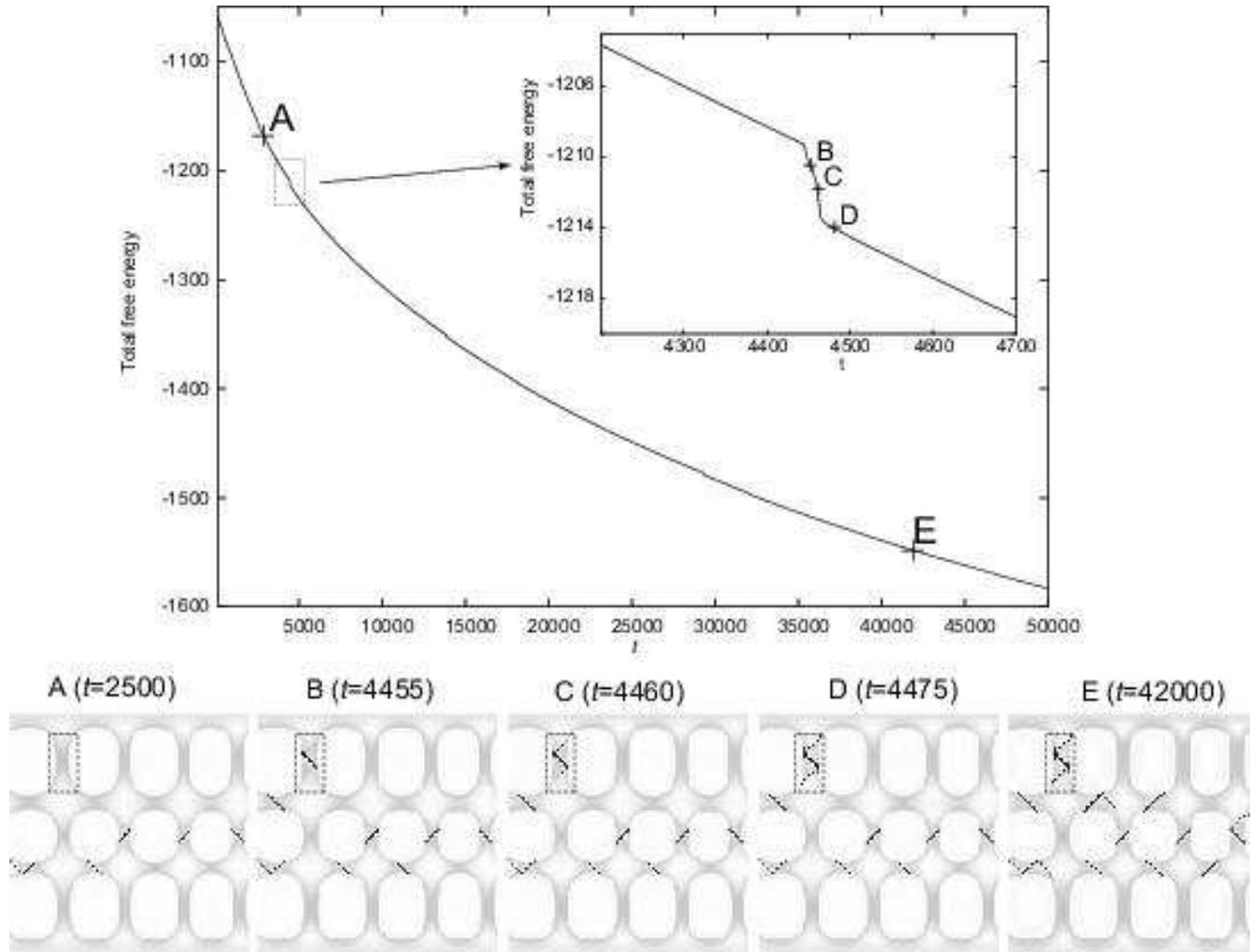}
\caption{
Relaxation of the total free energy 
$F$ in units of $\mu_{20}v_0= 20 k_{\rm B}T_0$ 
after a two-stepnch from 
$T/T_0= 2$ to 1 with 
the initial 
configuration in Fig.9. It 
mostly relaxes due to the gradual composition 
adjustment, but it sometimes relaxed 
due to dislocation formation 
as enlarged in the inset. Snapshots of $e$
 at the points A $\sim $ E  are given in 
the lower plates.
 }
\label{10}
\end{figure*}

\begin{figure}[t]
\includegraphics[scale=0.3]{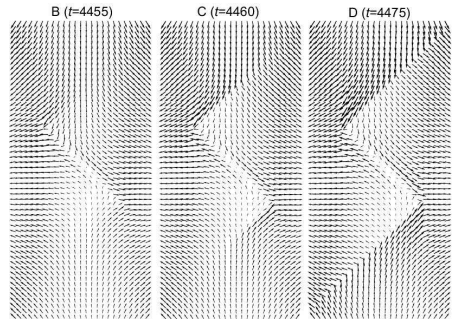}
\caption{
Elastic displacement $\bi u$  
in the marked regions 
 B,C, and D.}
\label{11}
\end{figure}

\begin{figure}[t]
\includegraphics[scale=0.8]{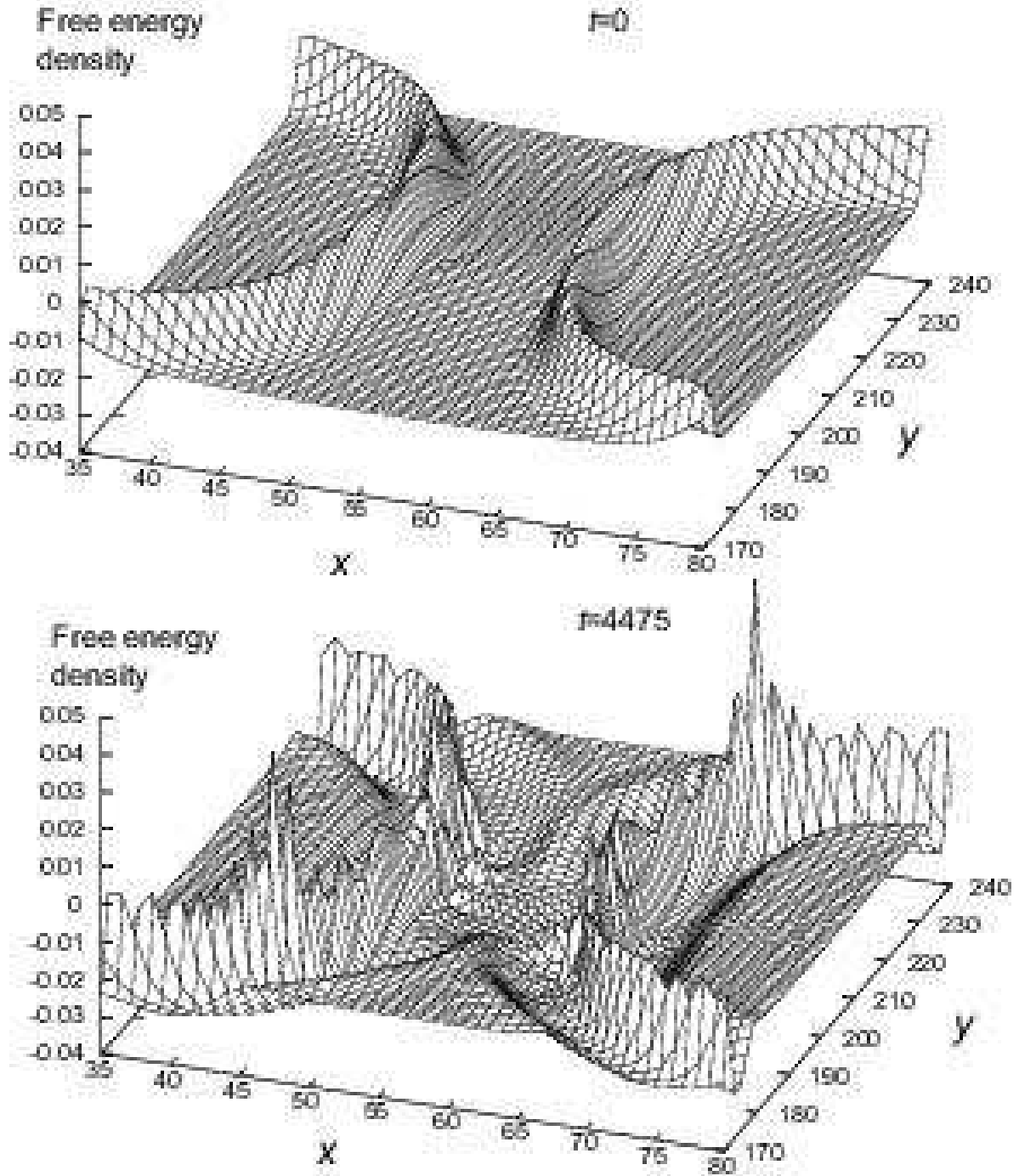}
\caption{
Bird views of the free energy 
 density $f$ in units of  $\mu_{20}
= 20 k_{\rm B}T_0v_0^{-1}$ 
at $t=0$ and 4475 after 
the two-step quench in Fig.10.}
\label{12}
\end{figure}

\begin{figure}[t]
\includegraphics[scale=0.7]{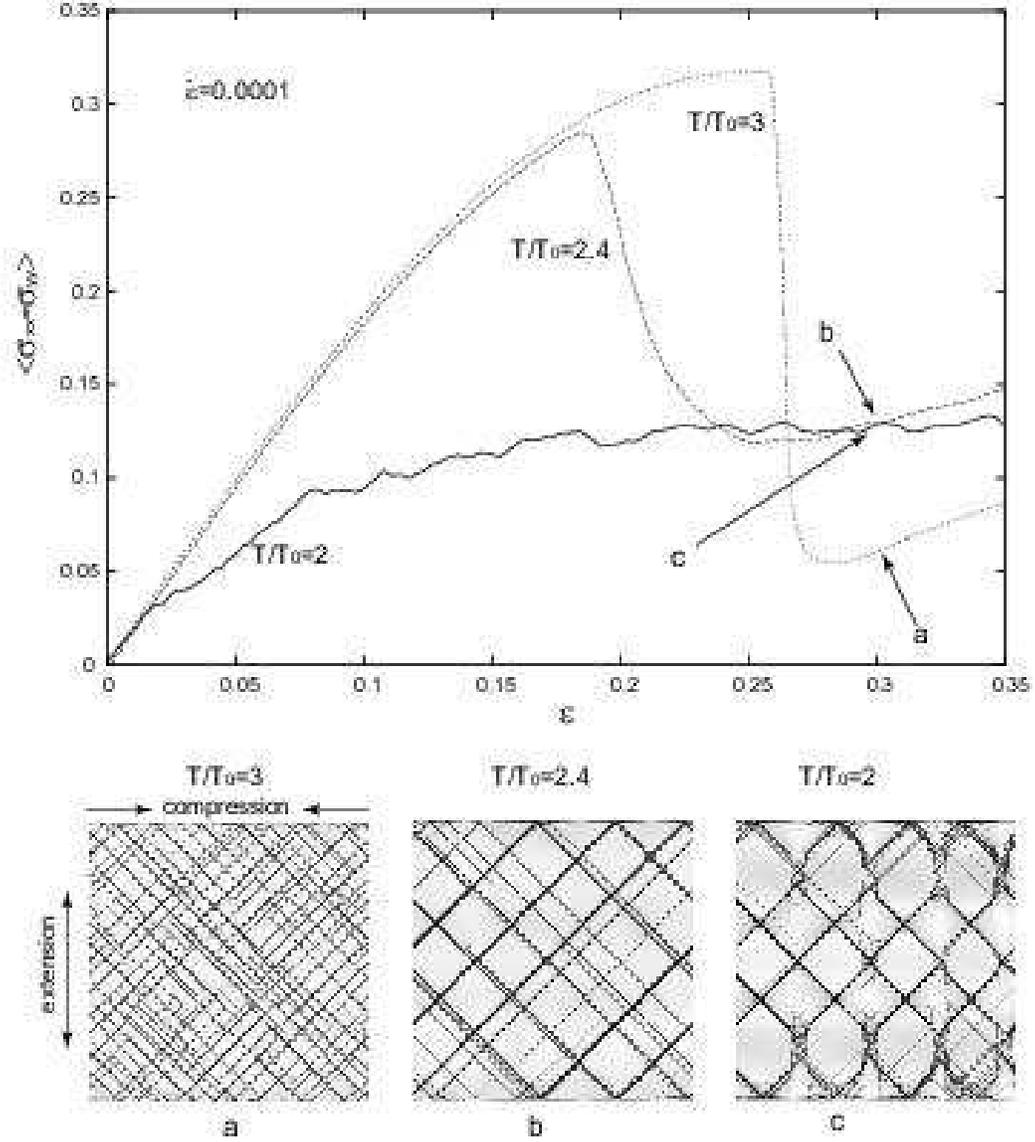}
\caption{
Stress-strain curves after application of uniaxial stretching 
$\epsilon= \dot{\epsilon}t$ with 
 $\dot{\epsilon}=10^{-4}$ for $T/T_0=3, 2.4$, and 2. 
There is  no dislocation at $t=0$. 
Snapshots of $e$ in (4.4) at points a,b, and c are given below, 
which represent slip patterns in  plastic flow. 
 }
\label{13}
\end{figure}

\begin{figure}[t]
\includegraphics[scale=0.7]{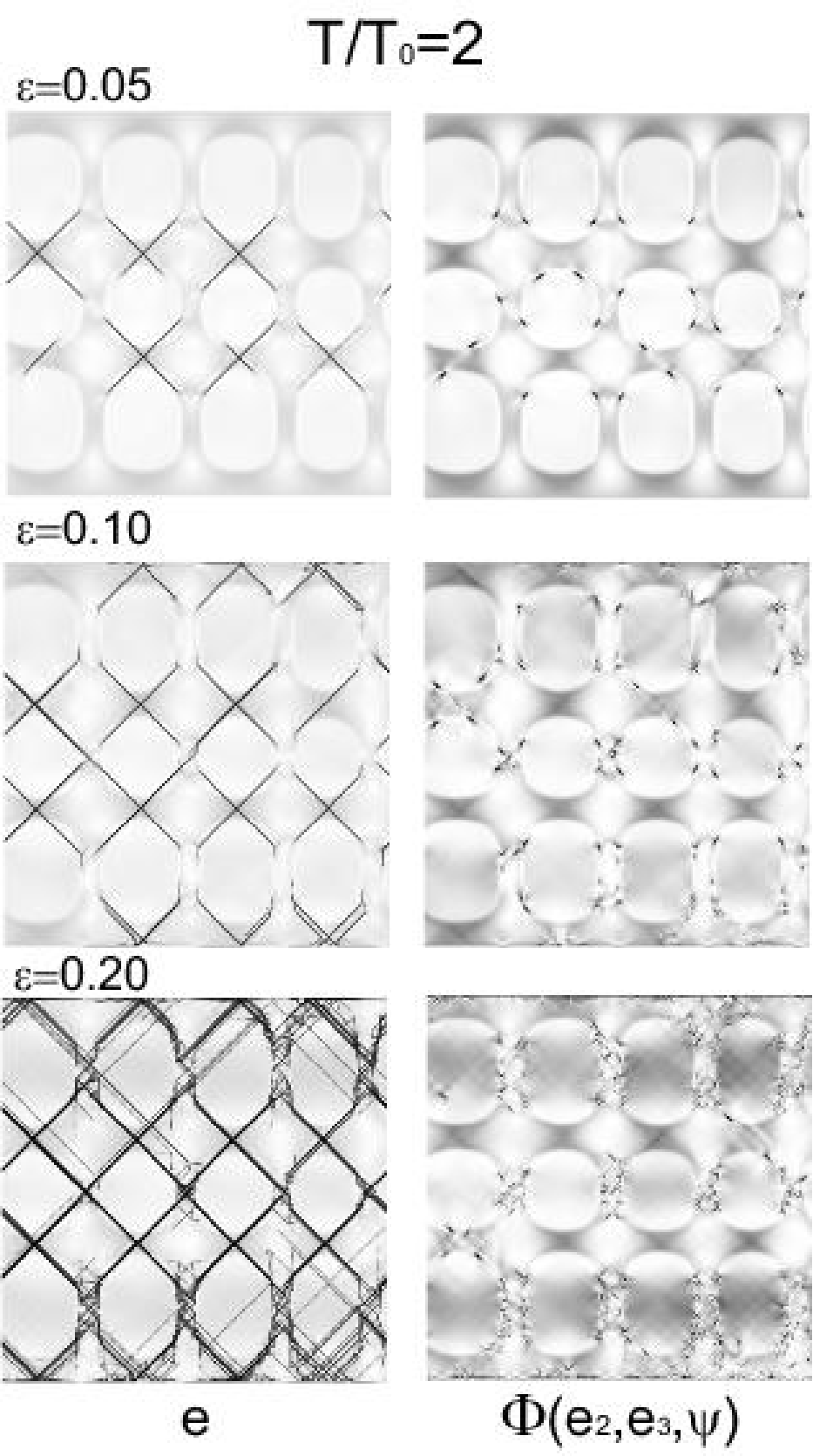}
\caption{
Snapshots of  $e$ in (4.4) 
and the shear deformation 
energy density $\Phi$ in (2.7) 
at $T/T_0=2$ for 
$\epsilon= 0.05, 0.1,$ and 0.2. 
 }
\label{14}
\end{figure}


\begin{references}
\bibitem{convection} 
In simulations with these terms 
the time mesh size $\Delta t$ must be 
made very small when dislocations are formed. 
Such attempts (in the uniaxial 
deformation case in Figs.13 and 14) 
did not essentially  alter 
the results.
\bibitem{comment2} 
The linear elasticity theory \cite{Landau} 
yields $e_1={\rm const.}y/(x^2+y^2)$ 
for a single edge dislocation at the origin. 
The expression for a slip is obtained in the presence of 
two dislocations  with opposite 
Burgers vectors. See Ref.$[10]$.
\end{references}
\end{document}